\documentclass{aa}
\usepackage{color}
\usepackage{graphicx}
\usepackage{txfonts}
\usepackage[colorlinks, citecolor=blue]{hyperref}
\usepackage{mathtools}
\usepackage{longtable} 
\usepackage{pdflscape} 
\usepackage{ulem}

\definecolor{gray}{RGB}{128,128,128}

\renewcommand{\vec}{\ensuremath\boldsymbol}
\newcommand{\mat}[1]{\ensuremath\boldsymbol{\mathsf{#1}}}

\begin{document}

   \title{Gaia GraL: {\it Gaia} DR2 Gravitational Lens Systems. III.\\ A systematic blind search for new lensed systems}
    
   \author{L. Delchambre\inst{1},
           A. Krone-Martins\inst{2},
           O. Wertz\inst{3},
           C. Ducourant\inst{4},
           L. Galluccio\inst{5},
           J. Kl\"{u}ter\inst{6},
           F. Mignard\inst{5},
           R. Teixeira\inst{7},
           S. G. Djorgovski\inst{8},
           D. Stern\inst{9},
           M. J. Graham\inst{8},
           J. Surdej\inst{1},
           U. Bastian\inst{6},
           J. Wambsganss\inst{6, 10},
           J.-F. Le Campion\inst{4},
           E. Slezak\inst{5}
        }

  \institute{
         Institut d'Astrophysique et de G\'{e}ophysique, Universit\'{e} de Li\`{e}ge, 19c, All\'{e}e du 6 Ao\^{u}t, B-4000 Li\`{e}ge, Belgium\\
          \email{ldelchambre@uliege.be}
         \and
         CENTRA, Faculdade de Ci\^encias, Universidade de Lisboa, 1749-016 Lisboa, Portugal
          \and
         Argelander-Institut f\"{ u}r Astronomie, Universit\"{ a}t Bonn,  Auf dem H\"{ u}gel 71, 53121 Bonn, Germany
          \and
          Laboratoire d'Astrophysique de Bordeaux, Univ. Bordeaux, CNRS, B18N, all{\'e}e Geoffroy Saint-Hilaire, 33615 Pessac, France
          \and
         Universit\'{e} C\^{o}te d'Azur, Observatoire de la C\^{o}te d'Azur, CNRS, Laboratoire Lagrange, Boulevard de l'Observatoire, CS 34229, 06304 Nice, France
          \and
        Zentrum f\"{u}r Astronomie der Universit\"{a}t Heidelberg, Astronomisches Rechen-Institut, M\"{o}nchhofstr. 12-14, 69120 Heidelberg, \\Germany
          \and
          Instituto de Astronomia, Geof\'isica e Ci\^encias Atmosf\'ericas, Universidade de S\~{a}o Paulo, Rua do Mat\~{a}o, 1226, Cidade Universit\'aria, 05508-900 S\~{a}o Paulo, SP, Brazil
        \and
         California Institute of Technology, 1200 E. California Blvd, Pasadena, CA 91125, USA
        \and
        Jet Propulsion Laboratory, California Institute of Technology, 4800, Oak Grove Drive, Pasadena, CA 91109, USA
        \and
        International Space Science Institute (ISSI), Hallerstra\ss e 6, 3012 Bern, Switzerland
          }

   \date{Received July 9, 2018; accepted ???, ???}

  \abstract
   {} 
   {In this work, we aim to provide a reliable list of gravitational lens candidates based on a search performed over the entire {\it Gaia} Data Release 2 ({\it Gaia} DR2). We also aim to show that the sole astrometric and photometric informations coming from the {\it Gaia} satellite yield sufficient insights for supervised learning methods to automatically identify strong gravitational lens candidates with an efficiency that is comparable to methods based on image processing.}
   {We simulated 106,623,188 lens systems composed of more than two images, based on a regular grid of parameters characterizing a non-singular isothermal ellipsoid lens model in the presence of an external shear. These simulations are used as an input for training and testing our supervised learning models consisting of Extremely Randomized Trees. The latter are finally used to assign to each of the 2,129,659 clusters of celestial objects extracted from the {\it Gaia} DR2 a discriminant value that reflects the ability of our simulations to match the observed relative positions and fluxes from each cluster. Once complemented with additional constraints, these discriminant values allowed us to identify strong gravitational lens candidates out of the list of clusters.}
   {We report the discovery of 15 new quadruply-imaged lens candidates with angular separations less than $6\arcsec$ and assess the performance of our approach by recovering 12 out of the 13 known quadruply-imaged systems with all their components detected in {\it Gaia} DR2 with a misclassification rate of fortuitous clusters of stars as lens systems that is below one percent. Similarly, the identification capability of our method regarding quadruply-imaged systems where three images are detected in {\it Gaia} DR2 is assessed by recovering 10 out of the 13 known quadruply-imaged systems having one of their constituting images discarded. The associated misclassification rate varying then between 5.83\% and 20\%, depending on the image we decided to remove.}
   {}
\titlerunning{Gaia GraL III -- Blind search for new lensed systems}
\authorrunning{L. Delchambre et al.}
\maketitle

\section{Introduction}
\label{sec:introduction}
    The last two decades have seen the advent of numerous large, deep sky and even time-resolved surveys such as the Two Micron All Sky Survey \citep[2MASS, ][]{Skrutskie2006}, the Catalina Real-Time Survey \citep[CRTS, ][]{2009ApJ...696..870D}, the {\it Wide-field Infrared Survey Explorer} \citep[{\it WISE}, ][]{Wright2010}, the Sloan Digital Sky Survey \citep[SDSS, ][]{Alam2015}, the Dark Energy Survey \citep[DES, ][]{DES2016}, the Panoramic Survey Telescope and Rapid Response System \citep[Pan-STARRS, ][]{Chambers2016}, the {\it Gaia} mission \citep{Gaia2016} and the Zwicky Transient Facility \citep[ZTF, ][]{2017NatAs...1E..71B}. Amongst these, {\it Gaia} stands out as the most accurate instrument performing a whole sky survey at the present time thanks to its impressive astrometric uncertainties at the $\mu$as level and excellent photometric sensitivity at the mmag level, down to a $G$ magnitude of 20.7 if isolated point-like sources are considered.

	Through all these surveys, hundreds of millions, to billions, of celestial objects are nowadays continuously observed over multiple wavelength ranges. This large amount of information, covering the whole celestial sphere, naturally yields to a greater chance of identifying rare objects such as $z > 7$ quasars \citep{Banados2018}, L and T sub-dwarf stars \citep[e.g.][]{Kirkpatrick1999,Kirkpatrick2014}, Type Ia supernovae \citep[e.g.][]{Jones2018} and multiply-imaged quasars \citep[e.g.][]{Inada2012, Agnello2017}.

	Strong gravitational lensing (hereafter GL) depicts the formation of multiple images of a background source whose light rays get deflected and distorted owing to the presence of a massive galaxy standing along the line-of-sight between the observer and the source. Although predicted by Einstein's gravitation theory \citep{Einstein1936}, it was not until 1979 that the first GL was finally discovered by \cite{Walsh1979}. Since then, GLs have found numerous applications in probing the nature of dark matter \citep{Dalal2002, Gilman2017}, in determining the shape of the dark matter halos of galaxies \citep{2018arXiv180709278S} or in clusters of galaxies \citep{Meneghetti2017, Jauzac2018}, as natural telescopes for detecting otherwise unobservable sources \citep{Peng2006,Zavala2018} or as a way to set constraints on cosmological parameters irrespective of the cosmic distance ladder \citep{Refsdal1964,Suyu2013,Tagore2018}. Notwithstanding their crucial importance in cosmology, the number of known GLs still remains very limited with just $\sim 200$ spectroscopically confirmed GLs out of which only $\sim 45$ are composed of more than two lensed images \citep[see e.g.][Table 1]{Ducourant2018}. Besides the fact that GLs are intrinsically rare, this scarcity is also due to the difficulty in identifying GLs in large astronomical catalogues.

	Conscious of the unique opportunity brought by these modern large sky surveys, numerous methods were recently developed to systematically search for GLs \citep{Bolton2008, More2016, Paraficz2016, Agnello2017, Jacobs2017, Lee2017, Agnello2018, Pourrahmani2018,2018arXiv181004480L}. At the state of the art, the Strong Gravitational Lens Finding Challenge \citep{Metcalf2018} is a recent effort to identify GLs in large scale imaging surveys as the upcoming Square Kilometer Array (SKA)\footnote{\href{http://skatelescope.org}{http://skatelescope.org}}, the Large Synoptic Survey Telescope (LSST)\footnote{\href{https://www.lsst.org}{https://www.lsst.org}} and the {\it Euclid} space telescope\footnote{\href{https://www.euclid-ec.org}{https://www.euclid-ec.org}}. The solutions envisaged to fulfill the proposed challenge encompass visual inspection procedures \citep{Hartley2017}, arcs and rings detection algorithms \citep{Alard2006, More2012, Sonnenfeld2018} and supervised machine learning methods \citep{Bertin2012, Avestruz2017, Lanusse2018}. Because GLs are rare objects, all these techniques have as a common objective to minimize the rate of contaminants among the predicted GL candidates.

	The {\it Gaia} space mission is mainly dedicated to the production of a dynamical three-dimensional map of our Galaxy \citep{Gaia2016}. In addition, it will provide valuable informations for millions of extragalactic objects \citep[e.g.][]{2012A&A...537A..42T, 2013A&A...556A.102K, 2013A&A...559A..74B, 2014A&A...568A.124D}, including the detection of new GLs \citep[e.g.][]{Agnello2017,2018arXiv181002624W,2018arXiv181004480L,Ostrovski2018}. Indeed, amongst the two billion objects that {\it Gaia} will observe, we expect $\sim2900$ GLs to be present in the final data release out of which more than 250 should have more than two lensed images \citep{Finet2016}. This constitutes an order of magnitude increase compared to the number of currently known GLs.

	In the present work, we aim to detect new GL candidates from a blind search performed over the entire {\it Gaia} DR2. To do so, we train and apply a supervised learning method, called Extremely Randomized Trees \citep[ERT, ][]{Geurts2006}, whose input are the precise relative positions and fluxes from clusters of celestial objects extracted out of the {\it Gaia} DR2. We concurrently show that these ERT models, despite their restricted input data (i.e. astrometry and photometry), can reach performances that are comparable to those of the best model from the Strong Gravitational Lens Finding Challenge \citep{Lanusse2018}. Specifically, we achieve a 90 percent identification rate of GLs with a misclassification rate of clusters of stars as GLs below one percent. A preliminary version of this method was already successfully used in \cite{KronMartin2018} in order to identify three GL candidates, out of which two were spectroscopically confirmed: GRAL113100-441959 by our group and WGD2038-4008 by \cite{Agnello2017} and independently selected by us. The present paper gives a detailed overview of the final method, of its performance and its application to the whole Gaia DR2.

    This study is carried out inside the {\it Gaia Gravitational Lenses} working group, or Gaia GraL, whose main objective is to unravel the possibilities offered by the ESA/{\it Gaia} satellite to identify and study gravitational lenses. This article is the third of a series of works produced based on the second data release of {\it Gaia} \citep[][hereafter {\it Gaia} DR2]{Gaia2018}.
	
	In Sect. \ref{sec:cluster_extraction} we present the methods we specifically developed for extracting clusters of objects out of the {\it Gaia} DR2. In Sect. \ref{subsec:simulated_gl}, we detail the use we made of the relative image positions and flux ratios of simulated GL systems so as to train supervised learning models with a view of identifying GL candidates out of the list of clusters (Sect. \ref{subsec:ert}). The performance of our classification algorithm is covered in Sect. \ref{subsec:performances}. Based on the resulting ERT predictions, a sample of the most promising GL candidates is given in Sect. \ref{subsec:gl_candidates}. Finally, we discuss our findings and conclude in Sect.\ref{sec:conclusion}. 
        

\section{Extraction of clusters of objects from Gaia DR2}
\label{sec:cluster_extraction}

\begin{figure*}
\begin{center}
\includegraphics[width=\textwidth]{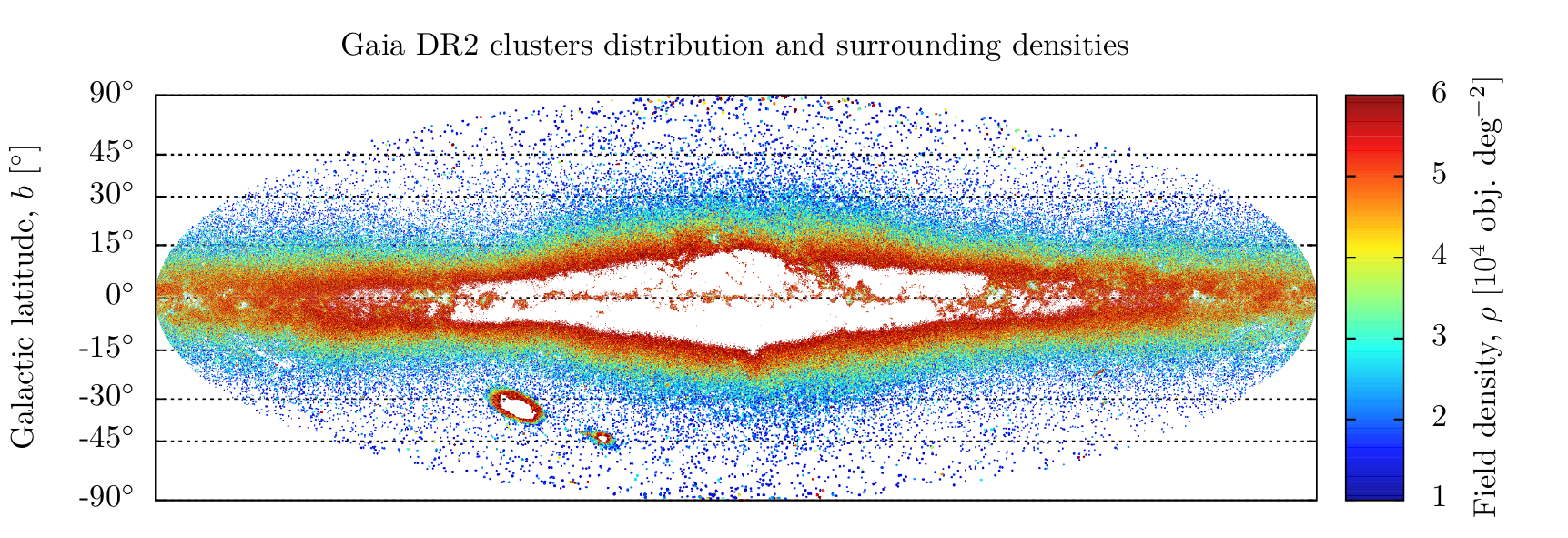}
\end{center}
\caption{Distribution of the 2,129,659 clusters of objects extracted from the {\it Gaia} DR2 catalogue. These are composed of three and four images that pass the soft astrometric test (see Section \ref{sec:cluster_extraction}), that have a maximal angular separation between components that is smaller than $6\arcsec$, that have absolute differences in $G$ magnitudes of $< 4$ mag and that stand in regions of the celestial sphere where the mean field density is lower than $6 \times 10^4$ objects deg$^{-2}$. Lower density regions near the galactic center can be explained by the filtering occurring in the on-board processing in order to prevent memory from saturating in such very dense regions of the sky \citep{Gaia2016}.}
\label{fig:cluster_distrib_densities}
\end{figure*}

	Our blind search for new GL candidates first consists in the extractions of clusters of objects out of the entire {\it Gaia} DR2. The latter can be accessed through the {\it Gaia} Archive bulk retrieval data facility\footnote{\href{http://cdn.gea.esac.esa.int/Gaia/gdr2/gaia\_source/csv/}{http://cdn.gea.esac.esa.int/Gaia/gdr2/gaia\_source/csv/}}. The details of this extraction is covered in the present section while the resulting catalogue of clusters obtained from {\it Gaia} DR2 sources is presented in Appendix \ref{sec:catalogue}.

	Prior to this extraction, we recall that all deflected sources from strong GLs consist of extragalactic sources. We thus expect the lensed images to have negligible parallaxes, $\varpi$, and proper motions, $(\mu_{\alpha^*}$, $\mu_{\delta})$ where $\mu_{\alpha^*} = \mu_{\alpha} \cos \delta$. We hence beforehand filtered the {\it Gaia} DR2 using the {\it soft astrometric test} described in \cite{Ducourant2018}. Specifically, we rejected observations having either $\varpi - 3 \sigma_\varpi \geq 4$ mas or $|\,\mu\,| - 3 \sigma_{\mu} \geq 4$ mas yr$^{-1}$ (where $\mu$ stands for $\mu_{\alpha^*}$ and $\mu_{\delta}$). Adopting a conservative approach, we did not discard the observations when $\varpi$, $\mu_{\alpha^*}$, $\mu_{\delta}$ or their associated uncertainties were missing. The number of sources we used was then reduced from 1,629,919,135 that are present in the original {\it Gaia} DR2 down to 1,217,192,458 that passed the soft astrometric test. We may note that two known GLs from Table \ref{tbl:known_gl} do not pass the soft astrometric test: DESJ0405-3308 (where image having {\it Gaia} source identifier = 4883180423151513088 has $\mu_{\alpha^*} = 16.44 \pm 1.723$ mas yr$^{-1}$ and $\mu_{\delta} = -13.43 \pm 2.143$ mas yr$^{-1}$) and RXJ0911+0551 (where image having {\it Gaia} source identifier = 580537092879166848 has $\mu_{\alpha^*} = -7.76 \pm 1.026$ mas yr$^{-1}$). The very large proper motions observed in DESJ0405-3308 can be presumably explained by the fact that this image has a large astrometric excess noise of $\epsilon = 5.791$ mas, and is hence not astrometrically well-behaved.

	Because an exhaustive analysis of all combinations of objects from {\it Gaia} DR2 is infeasible and not desirable, we restricted our search for clusters to those having a finite angular size and a limited absolute difference in magnitude between their components. Extraction criteria were accordingly based on the characteristics of known quadruply-imaged systems from Table \ref{tbl:known_gl}. Amongst the listed GLs, all have angular sizes smaller than $5.8\arcsec$, with the exception of SDSS1004+4112. Also, none of them is composed of images having an absolute difference in $G$ magnitude, $\Delta G$, larger than 3.5 mag. Considering that the extraction of clusters comparable in size to SDSS1004+4112 (i.e. $\sim 15\arcsec$) would result in a too large fraction of fortuitous aggregations of stars in our final list of GL candidates, we finally adopted the following convention: clusters of celestial objects must have (i) three or four images in order to provide a sufficient number of constraints for identifying GL candidates out of these clusters as well as to enable their subsequent modeling, (ii) a maximal angular separation between any pair of images that is below $6\arcsec$, (iii) an absolute difference in $G$ magnitude between components lower than $4$ mag.

	Without any further constraints, we expect the vast majority of our GL candidates to naturally fall in regions of high stellar densities such as in the Galactic plane, in the Magellanic clouds or in globular clusters. In order to identify these high density regions, we evaluated the local density of objects around each cluster based on the Gaia DR2. A mean density of objects was accordingly computed within a radius of $30\arcsec$ around each cluster. This radius was selected as a trade-off between its statistical significance and its locality property. From Table \ref{tbl:known_gl}, nearly all known GLs reside in regions with a mean field density $\rho < 3 \times 10^4$ objects deg$^{-2}$, therefore avoiding globular clusters and dense regions of the Galactic plane (see Figure \ref{fig:cluster_distrib_densities}). None of the known quadruply-imaged GLs, with the exception of J2145+6345 that was discovered after the submission of the present paper \citep{2018arXiv181004480L}, reside in regions with mean field density $\rho \geq 6 \times 10^4$ objects deg$^{-2}$. Accordingly, we set an upper limit on the density of objects surrounding each cluster of $6 \times 10^4$ objects deg$^{-2}$.
	
    The production of the list of clusters is based on a search for neighbors around each of the {\it Gaia} DR2 sources that passed the soft astrometric test\footnote{We used for this purpose a subdivision of the celestial sphere based on the Hierarchical Triangular Mesh \citep{Kunszt2001}.}. All combinations of three or four images were considered to produce the final list of clusters as the deflecting galaxy or contaminating stars might be present within the identified clusters.
    
    Each cluster was then assigned a unique identifier, which is based on the mean position of its constituent sources. The common convention of taking the position of the brightest image as an identifier for GL systems was not adopted here as it would lead to ambiguities in identifying clusters for which all combinations of images were explored. Figure \ref{fig:cluster_distrib_densities} illustrates the distribution of the 2,129,659 clusters derived from {\it Gaia} DR2, amongst which 2,058,962 are composed of three components and 70,697 are composed of four components. Also depicted are the mean field densities that are associated with each of these clusters.

\section{Identification of the lens candidates from supervised learning method}
\label{sec:method}

\begin{table*}
\caption{List of all spectroscopically confirmed quadruply-imaged systems having $N_{\mathrm{img}} = {3, 4}$ components detected in {\it Gaia} DR2. The lens size and the maximal absolute difference in magnitude and color, $\Delta G$ and $\Delta (G_{\mathrm{BP}}-G_{\mathrm{RP}})$, are computed over all combinations of the lensed images while the field density is estimated within a radius of $30\arcsec$ around the system. Numbers in parentheses correspond to the number of images that were used in the computation of the maximal absolute difference in color. Four image configurations are processed using the ERT model ABCD as well as using all the ABC, ABD, ACD and BCD models based on the appropriate combination of images (see Section \ref{sec:method}). Three image configurations are processed using models ABC, ACD, ABD and BCD.}
\footnotesize
\label{tbl:known_gl}
\begin{center}
\begin{tabular}{lrrrrrrrrrr}
\hline
\hline
Lens identifier & $N_{\mathrm{img}}$ &  Size & $\Delta G$ & $\Delta (G_{\mathrm{BP}}-G_{\mathrm{RP}})$ & Field density & \multicolumn{5}{c}{ERT probabilities}  \\
 & & [mas] & [mag] & [mag] & [obj. deg$^{-2}$] & ABCD & ABC & ABD & ACD & BCD \\
\hline
2MASSJ11344050-2103230 & 4 & 3682 & 1.76 & 0.16(3) & 18335 & 1.00 & 0.97 & 0.66 & 0.72 & 0.31 \\
J1606-2333 & 4 & 1723 & 0.76 & 0.07(2) & 50420 & 1.00 & 0.9 & 0.99 & 0.94 & 0.84  \\
WGD2038-4008 & 4 & 2869 & 0.65 & & 22918 & 1.00 & 1.00 & 0.86 & 0.7 & 0.72  \\
HE0435-1223 & 4 & 2539 & 0.73 & 0.29(2) & 18335 & 0.99 & 0.92 & 0.7 & 0.75 & 0.78 \\
PG1115+080 & 4 & 2427 & 1.79 & 0.02(3) & 18335 & 0.99 & 0.98 & 0.99 & 0.75 & 0.69 \\
B1422+231 & 4 & 1281 & 3.46 & & 22918 & 0.98 & 0.87 & 0.98 & 0.83 & 0.59 \\
J2145+6345\;$^{(1)}$\;$^{(2)}$ & 4 & 2068 & 1.70 & & 73339 & 0.97 & 0.96 & 0.92 & 0.95 & 0.86 \\
2MASXJ01471020+4630433 & 4 & 3262 & 2.37 & 0.10(2) & 41253 & 0.95 & 0.66 & 0.99 & 0.55 & 0.51 \\
2MASSJ13102005-1714579 & 4 & 5735 & 1.21 & 0.21(3) & 22918 & 0.92 & 0.93 & 0.92 & 0.85 & 0.84  \\
J1721+8842\;$^{(3)}$ & 4 & 3906 & 1.75 & 0.13(2) & 27502 & 0.91 & 0.32 & 0.99 & 0.65  & 0.84  \\
WFI2033-4723 & 4 & 2533 & 1.18 & 0.04(2) & 18335 & 0.89 & 0.98 & 0.8 & 0.67 & 0.99 \\
SDSS1004+4112\;$^{(2)}$ & 4 & 14732 & 1.35 & 0.22(4) & 22846 & 0.88 & 0.99 & 0.24 & 0.99 & 0.19  \\
RXJ1131-1231 & 4 & 3232 & 2.11 & & 22918 & 0.65 & 0.69 & 0.93 & 0.56 & 0.72 \\
SDSSJ1433+6007 & 3 & 3754 & 0.38 & 0.04(2) & 13751 & & 0.99 & 0.92 & 0.52 & 0.65  \\
DESJ0405-3308\;$^{(1)}$\;$^{(2)}$ & 3 & 1416 & 0.20 & 0.32(3) & 13751 & & 0.98 & 0.78 & 0.66 & 0.93 \\
J0408-5354 & 3 & 4184 & 1.08 & 0.30(2) & 22918 & & 0.42 & 0.56 & 0.97 &  0.95 \\
HE0230-2130 & 3 & 2188 & 1.03 & & 13751 & & 0.95 & 0.75 & 0.08 & 0.05 \\
H1413+117 & 3 & 1111 & 0.25 & 0.12(2) & 18335 & & 0.86 & 0.48 & 0.6 & 0.65 \\
RXJ0911+0551$^{(2)}$ & 3 & 3260 & 1.11 & 0.04(2) & 27502 & & 0.65 & 0.19 & 0.67 & 0.07 \\
J0630-1201 & 3 & 1901 & 0.11 & 0.33(3) & 36669 & & 0.58 & 0.1 & 0.09 & 0.3 \\
WISE025942.9-163543 & 3 & 1577 & 0.76 & & 13751 & & 0.32 & 0.45 & 0.16 & 0.32 \\
\hline
\multicolumn{11}{p{0.9\textwidth}}{{\protect\textbf{Notes}\hspace{0.5em}1.} J2145+6345 and DESJ0405-3308 are not contained in the original list of known GLs from \cite{Ducourant2018} as their discovery was announced after the list was actually compiled \citep[see][respectively]{2018arXiv181004480L,2018MNRAS.480.5017A}.} \\
\multicolumn{11}{p{0.9\textwidth}}{{\protect\textbf{Notes}\hspace{0.5em}2.} J2145+6345, SDSS1004+4112, DESJ0405-3308 and RXJ0911+0551 are not part of the final catalogue of clusters (Appendix \ref{sec:catalogue}) because they are standing in regions of high stellar density (J2145+6345), because one of their images do not pass the soft astrometric test (DESJ0405-3308 and RXJ0911+0551) or because of they have a too large angular size (SDSS1004+4112).} \\
\multicolumn{11}{p{0.9\textwidth}}{{\protect\textbf{Notes}\hspace{0.5em}3.} Regarding J1721+8842, we selected the configuration of four images out of the five images available for which the ERT probabilities was the highest (0.91). Alternative combinations implying image with {\it Gaia} source identifier = 1729026466114871296 provide ERT probabilities between 0 and 0.05.}
\end{tabular}
\end{center}
\end{table*}

\begin{figure*}
\includegraphics[width=\textwidth]{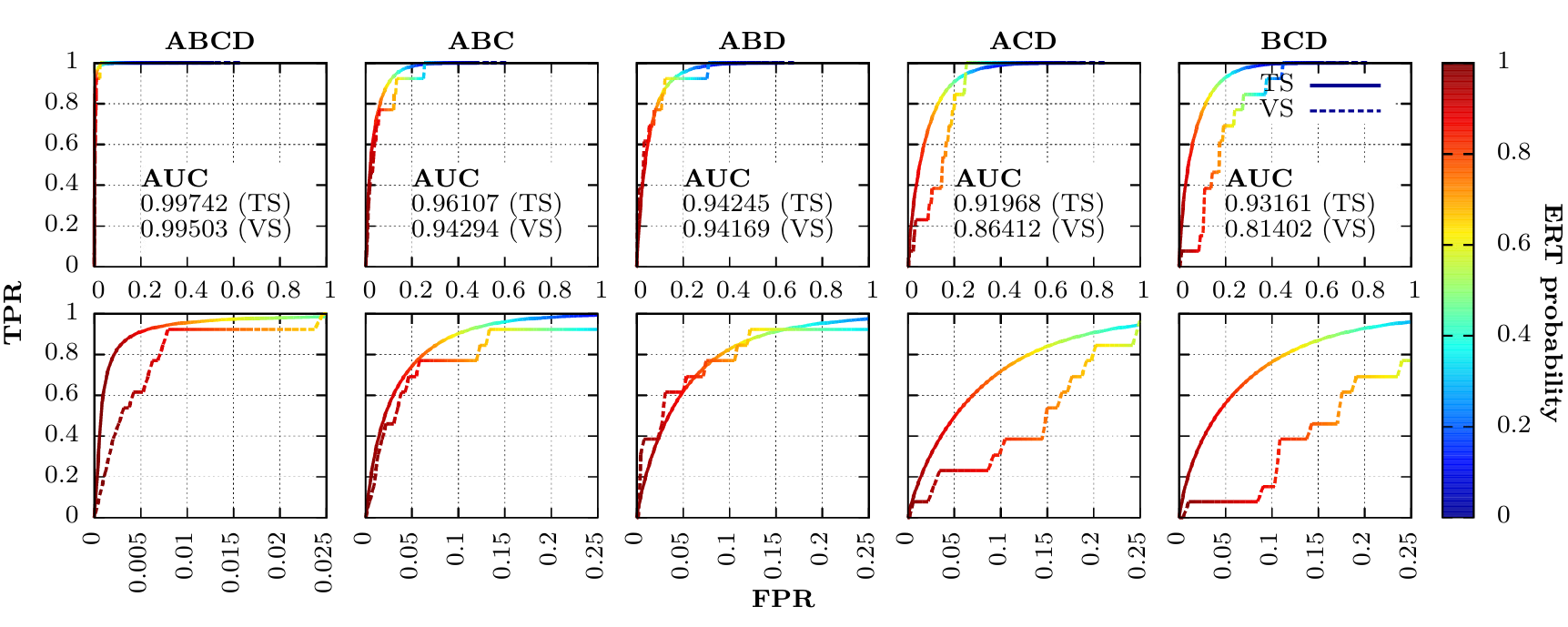}
\caption{Receiver Operating Characteristics (ROC) curve of the ABCD, ABC, ABD, ACD and BCD models based on the test set (TS) and validation set (VS) of observations. Upper panels show the entire ROC curves whereas lower panels concentrate on the low FPR regions of each curve. The classification performances of each model is evaluated through the computation of the area under each of the TS- and VS--ROC curves (AUC).}
\label{fig:roc}
\end{figure*}

	After we defined the list of clusters, the second step in our methodology to perform a blind search for new GLs was the assignment of a discriminating value, called the Extremely Randomized Trees (ERT) probability, to each cluster. These ERT probabilities, $P$, do not constitute probabilities in a mathematical sense and should rather be viewed as a figure-of-merit or as a score that reflects the ability for the clusters to be matched to the image positions and relative magnitudes produced from simulations of lens systems (see Section \ref{subsec:simulated_gl}). They can however be translated into expected ratios of identification of GLs and to expected ratios of misclassification of groups of stars as GLs through the use of an appropriate cross-validation procedure.

\subsection{Catalogue of simulated gravitational lenses}
\label{subsec:simulated_gl}
	Supervised learning methods aim to automatically discover the relations that may exist between a set of input attributes and an associated outcome value based on a collection of training instances. The more complete and representative the learning set of observations, the fairer and more accurate are the resulting predictions \citep[e.g.][]{2017MNRAS.468.4323B}. Training sets can be constructed either directly by using observational data or by using simulations. Regarding the specific problem of the identification of GL candidates, the limited number of 44 known quadruply-imaged systems from the list of \cite{Ducourant2018} out of which only 19 have more than two images that are detected in {\it Gaia} DR2 forces us to rely on simulations to cover the input space of attributes in a satisfying manner.
    
	To construct our training set, we consider a non-singular isothermal ellipsoid lens model in the presence of an external shear \citep[hereafter NSIEg lens model, ][see Appendix \ref{sec:nsieg} for a concise description]{Kormann1994, Kovner1987}. This model has been proven to be well suited for reproducing the relative positions and flux ratios of the lensed images when the deflector is a massive late-type galaxy. A consequence of choosing this specific model is that the GL candidates we will identify through supervised learning methods will be restricted to systems which can be well modeled by this particular model. However, this does not constitute a major drawback to our implementation as most of the known lens systems are generally well described by this particular model.

Accordingly, we simulated 106,623,188 GL systems having four observable images based on a plausible range of parameters for the NSIEg lens model as listed in Table \ref{tbl:nsieg_parameter_range}. 
For completeness, we note that a redundancy exists amongst the produced simulated GL systems. This is a natural consequence of a NSIEg lens model as, for example, all lens models with a shear orientation of $\pi - \omega$ and source position $(-x_s, y_s)$ are the horizontal reflections of models with a shear orientation of $\omega$ and a source position of $(x_s, y_s)$.

\begin{table}
\footnotesize
\caption{Range of parameters explored for producing the simulated lens catalogues from a NSIEg lens model. A detailed description of these parameters is provided in Appendix \ref{sec:nsieg}.}
\label{tbl:nsieg_parameter_range}
\begin{center}
\begin{tabular}{lrrrrrrr}
\hline
\hline
Explored & \multicolumn{6}{c}{NSIEg parameters} \\
range & $b$ & $q$  & $s$  & $\gamma$ & $\omega$   & $x_s$, $y_s$ \\
\hline
Range start     & 1   & 0.2  & 0    &  0       & 0$\degr$   & - 1 \\
Step            &  -- & 0.05 & 0.01 & 0.01     & 5$\degr$   & 0.02 \\
Range end       &  -- & 1    & 0.3  & 0.3      & 175$\degr$ & 1 \\
\hline
\end{tabular}
\end{center}
\end{table}

\subsection{The Extremely Randomized Trees supervised learning models}
\label{subsec:ert}
ERT probabilities are derived from a tree-based machine learning algorithm that relies on the assumption that the aggregation of the predictions of several weak, strongly randomized trees can compete or even surpass more sophisticated methods like artificial neural networks \citep{Haykin1998} or support vector machines \citep{Cortes1995}. This assumption was initially supported by \cite{Mingers1989} and was later successfully implemented in the Boostrap aggregating meta-algorithm \citep{Breiman1996} as well as Random Forests \citep{Breiman2001}. Whereas the usual classification and regression trees (CART) select at each node of the tree the attribute and cut-value within this attribute that best split the learning set of observations associated with this node according to a given score measure (e.g. the reduction of the variance in regression problems or the information gain in classification problems), the ERT algorithm instead picks up a subset of $K$ random attributes as well as a random cut-point within each of these attributes so as to select the split that maximizes the given score measure \citep{Geurts2006}. The algorithm ends when no more than $n_{\mathrm{min}}$ training observations remain in all leaf nodes. The aggregation of the predictions of $N$ trees (a majority vote in our case) then lessens the variance of the ERT, in the sense that it avoids the model to be too specific to the learning set of observation we used while not being able to generalize the learned relations over unseen observations.

	As we expect some of the lensed images to be missing from {\it Gaia} DR2 (see Table \ref{tbl:known_gl} for examples), all combinations of three and four images were considered for building the ERT. Also, as the central and strongly de-magnified image produced in NSIEg-like GL is often out of reach of the {\it Gaia} photometric sensitivity, it was not taken into account. These ERT models will be referred to in the following as ABCD, ABC, ABD, ACD and BCD where A, B, C, D identify the images we used during the learning phase of the corresponding ERT, assuming these are sorted in ascending order of $G$ magnitude.

	All ERT models were trained using a {\it learning set} of observations (LS) composed of half the number of simulations we described in Section \ref{subsec:simulated_gl}, plus an equal number of contaminant observations for which both the image positions and magnitudes were randomly drawn from a uniform distribution. Note that these simulated contaminants are still restricted to have an absolute difference in magnitude, $\Delta G$, lower than 4 mag, in agreement with the requirements we developed in Section \ref{sec:cluster_extraction}. The other half of the simulations is then kept as a {\it test set} of observations (TS) for cross-validation purpose, after being complemented by the addition of simulated contaminants. We should stress here that neither LS, nor TS follow a realistic distribution of positions and magnitudes as it would require, for example, the distribution of the eccentricity of the lensing galaxy or the properties of clusters of stars to be taken into account. These were solely designed with the aim of identifying the regions of the input space of parameters (i.e. relative positions and fluxes) where GLs and contaminants are situated through the use of the ERT.
    
    We then added a Gaussian noise to the images positions, $\sigma_{xy}$, and magnitudes, $\sigma_G$, for each of the learning set and test set configurations, before discarding some of their images in order to create the input instances used in the ABCD, ABC, ABD, ACD and BCD models. These configurations are then normalized through an orthogonal transformation and a scaling to have their brightest image (image A) repositioned at $\left(0,0\right)$ along with a magnitude of $0$ and their faintest image (image C or D, depending on the number of images of the configuration) repositioned at $\left(1,0\right)$.

    The addition of noise to the simulations in the present case is not designed to take into account the astrometric and photometric uncertainties of {\it Gaia} DR2, which are actually much lower than the noise we introduced. Rather, this noise was added to deal with the possible imperfections of the NSIEg model, and to enable the machine learning method to deal with lenses that depart from this idealized model (e.g. due to sub-structures in the deflecting galaxies or to the inherent fact that the NSIEg lens model only constitutes an approximation of GLs whose deflectors are late-type galaxies). Similarly, the noise added to the magnitudes reflect the fact that micro-lensing events frequently occur due to galaxy sub-structures. Also, because of the difference in the optical paths of the lensed images, time-delays exist between each of them, such that if the deflected source is a variable object, like quasars are, discrepancies would exist between our instantaneous noiseless simulations and real observations.

	Regarding the ERT model ABCD, the set of input attributes is composed of the normalized images positions, $\left(x_B, y_B\right)$ and $\left(x_C, y_C\right)$, of the normalized $G$ magnitudes $G_B, G_C, G_D$ and of their respective differences $\left(x_B-x_C, y_B-y_C\right)$, $G_C-G_B$, $G_D-G_B$ and $G_D-G_C$. We remind that, because of the normalization procedure,  $x_A = y_A = y_D = G_A = 0$ while $x_D = 1$. The attributes used in the ERT model ABC is then similarly given by $(x_B, y_B)$, $G_B$, $G_C$ and $G_C-G_B$, from which the attributes used in the ABD, ACD and BCD models can be easily extrapolated.

	The parameters of the ERT models (i.e. $N$, $K$ and $n_{\mathrm{min}}$) as well as the level of noise we add to each of the LS and TS configurations, $\sigma_{xy}$ and $\sigma_G$, were chosen in a heuristic way based on the identification performance of a {\it validation set} of observations (VS). The latter is composed of the known lensed systems having four detections in Gaia DR2\footnote{J2145+6345 was not used for building our ERT models, nor for determining the level of noise to add to our simulations as this lens was not already published at the time of submission.}, as listed in Table \ref{tbl:known_gl}, and of $10^6$ clusters we randomly extracted from {\it Gaia} DR2 with a size smaller than $30\arcsec$ and $\Delta G < 4$ mag. Various combinations of these parameters were probed in the ranges $N \in \left[10, 1000\right]$, $K \in \left[2, 12\right]$, $n_{\mathrm{min}} \in \left[2, 8\right]$, $\sigma_G \in \left[0, 1\right]$ and $\sigma_{\mathrm{xy}} \in \left[0, 0.1\,s\right]$ where $s$ stands for the lens size. The set of parameters we selected regarding the ABCD models: $N = 100$, $K = 12$, $n_{\mathrm{min}} = 2$, $\sigma_{xy} = 0.01 s$ and $\sigma_{G} = 0.25$, yield to a satisfactory fraction of 75\% of GLs that are correctly identified along with a misclassification rate of clusters of stars as GLs of 0.625\% if $P > 0.9$. Without much surprise, tests performed on the ABC, ABD, ACD and BCD models lead to the same set of parameters, at the exception of $K = 5$, though the associated identification capabilities are now getting largely hampered.

\subsection{Performances on the identification of known and simulated gravitational lenses}
\label{subsec:performances}
	
	The performances of each model in classifying GL candidates were assessed by computing, for each of them, the fraction of GLs that are correctly identified (the true positive rate) and the fraction of clusters of stars that are misclassified as GLs (the false positive rate). By reporting the true and false positive rates (hereafter TPR and FPR, respectively) that are associated with all ERT probabilities in a graph, we obtain the so-called {\it Receiver Operating Characteristics} (ROC) curves, shown in Figure \ref{fig:roc}. In the latter, the area under the ROC curve (AUC) is a commonly used measure of the classification capability of each model. For completeness, we have to note that some simulated contaminants from our training set can not be differentiated from the relative image positions and fluxes produced through NSIEg lens models. Regarding our ERT models, this has the effect of decreasing the TPR evaluated on the test set while increasing the associated FPR. Still, as this degeneracy is seen in real observations, we decided not to remove these ambiguous simulations from our training set.

	We can see from Figure \ref{fig:roc} and Table \ref{tbl:known_gl} that the ERT model ABCD is able to identify 12 out of the 13 known GLs (i.e. 92.3\%) and 92.5\% of the simulated GLs along with a FPR that is below one per cent if $P > 0.84$. The associated AUC (0.99503 if evaluated on the validation set or 0.99742 if evaluated on the test set) can be compared with the 0.98 obtained by the best lens classifier of the Strong Gravitational Lens Finding Challenge \citep{Lanusse2018} where a FPR of one per cent is associated with a TPR of 90\% \citep{Metcalf2018}. These numbers should however be taken with care given the difficulties in equitably comparing two models designed for different instruments, having different angular resolutions, photometric sensitivities and working directly on images, on one side and on reduced data, on the other side. In a more recent work, \cite{2018arXiv180806151W} achieve a TPR of 80\% along with a FPR of 2\% by directly modelling the quadruple lens systems through the fit of a right hyperbola to the observed relative positions of the lensed images \citep{1996ApJ...472L...1W}. Cuts on the resulting axis ratio, $q$, and on the scatter of the observed images around the fitted hyperbola being then used to select GL candidates. These comparisons demonstrate the efficiency of the approach adopted in the present work and more particularly of the ERT on this particular problem. These results also demonstrate the huge potential of {\it Gaia} regarding the identification of GLs, mostly coming from its impressive astrometric and photometric precision.

	Regarding the ERT models ABC and ABD, these provide FPRs of 5.83\% and of 7.48\%, respectively, on the validation set if these are associated with a TPR of 75\%. The FPR associated with the test set are 5.08\% and of 7.74\%, respectively, for the same TPR. Nevertheless, if a TPR of 75\% is considered for the models ACD and BCD then the corresponding FPR computed on the validation set rises to $\sim 20$\% ($\sim 10$\% on the test set). These larger FPRs apparently come from the intrinsic difficulty that the ERT have to identify GLs for which the two brightest images are not seen together, as illustrated by the ROC curves computed on the test set. Also, the discrepancies noticed in the ROC curves computed based on the test set, on one side, and on the validation set, on the other side, can be explained by the statistical fluctuations owing to the low number of known GLs present in the validation set (13) and by the fact that these contain different populations of lenses (i.e. the validation set contains a realistic population of lenses whereas the test set contains a population of simulated lenses coming from a uniform coverage of the NSIEg parameters). We note that FPRs as low as a few per cent still lead to a large number of contaminant observations in the final catalogue owing to the $2 \times 10^6$ clusters identified in {\it Gaia} DR2. The appropriate filtering of these numerous contaminants is described in Section \ref{subsec:gl_candidates}.
	
	Beside the overall performance of our approach, some of the known lenses from Table \ref{tbl:known_gl} are still being assigned low ERT probabilities, $P$, namely: J0630-1201 ($P_{\mathrm{ABC}} = 0.58$), RXJ0911+0551 ($P_{\mathrm{ACD}} = 0.67$), RXJ1131-1231 ($P_{\mathrm{ABCD}} = 0.65$) and WISE025942.9-163543 ($P_{\mathrm{ABD}} = 0.45$). The first of these, J0630-1201, is a recently discovered GL composed of five lensed images and two deflecting galaxies \citep{Ostrovski2018}, which can hence not be reproduced through a single NSIEg lens model. RXJ0911+0551 and RXJ1131-1231 have flux ratios that are difficult to reproduce without including microlensing by small-scale structures in the lens galaxy \citep{Keeton2003}. The fact that RXJ1131-1231 obtains an ERT probability of $P_{ABD} = 0.93$ once image C is discarded also supports this hypothesis. The study of the recently discovered GL WISE025942.9-163543 currently remains very limited, though the preliminary modeling performed by \cite{Schechter2018} using a non-singular isothermal sphere lens model in the presence of external shear (i.e. NSIEg lens model with $q = 1$ and $s = 0$) already highlighted the difficulties in reproducing the observed flux ratios,even if the image positions can be fairly well reproduced by this kind of model \citep{2018arXiv180806151W}. The modeling that we have performed using a NSIEg lens model has led to the same conclusion.
	
	We also note that two GL candidates, PS1J205143-111444 and WGD2141-4629, were already present in the original list of \cite{Ducourant2018}. These obtain maximal ERT probabilities of $P_{\mathrm{ABD}} = 0.01$ and of $P_{\mathrm{ABC}} = 0.62$, respectively. Whereas PS1J205143-111444 is probably not a GL that is reproducible through a NSIEg lens model, the lensing nature of WGD2141-4629 remains uncertain though highly improbable because of the non-negligible, opposite proper motions of two of its images while one of them is astrometrically well-behaved (i.e. astrometric excess noise of $\epsilon = 0$ mas). More recently, \cite{2018arXiv180511103A} discovered two new quadruply-imaged lens candidates, WG210014.9-445206.4 and WG021416.37-210535.3, that are not part of our input list of candidate lenses taken from \cite{Ducourant2018}. These candidates obtain maximal ERT probabilities of $P_{\mathrm{ABC}} = 0.4$ and $P_{\mathrm{ABC}} = 0.94$, respectively. Despite the fact that the lensing nature of WG210014.9-445206.4 looks very promising, it was not recognized by our mehod. Possible reasons are the finite identification rate (TPR) of the ERT or the hardly reproducible relative positions and magnitudes of this system through a NSIEg lens model.

	Finally, we mention that the ERT models described here differ significantly from those we built and used in Paper I \citep{KronMartin2018} as we adjusted our model to known GLs contained in {\it Gaia} DR2, whereas only SDSS J1004+4112 had all its components detected in {\it Gaia} DR1.

\subsection{Identification of new gravitational lens candidates}
\label{subsec:gl_candidates}

\begin{table*}
\caption{List of GL candidates. The finding charts depicting all of these candidates are given in Figure \ref{fig:finding_charts}. Numbers in parentheses correspond to the number of images that were used in the computation of the maximal absolute difference in color, $\Delta (G_{\mathrm{BP}}-G_{\mathrm{RP}})$.}
\label{tbl:gl_promising}
\footnotesize
\begin{center}
\begin{tabular}{rlrrrrrrrr}
\hline
\hline
Num. & Candidate identifier & $N_{\mathrm{img}}$ & Right ascension & Declination & Size & Field density & $\Delta G$ & $\Delta (G_{\mathrm{BP}}-G_{\mathrm{RP}})$ & ERT prob. \\
 & &  & [$\degr$] & [$\degr$] & [mas] & [obj. deg$^{-2}$] & [mag] & [mag] & \\
\hline
{[4]} & 214110146+314107480 & 4 & 325.292262 & 31.685426 & 3602 & 27502 & 0.67 &   & 1.00 \\
{[8]} & 053036992-373011003 & 3 & 82.654147 & -37.503067 & 1036 & 27502 & 2.99 &   & 0.98 \\
{[11]} & 153725327-301017053 & 3 & 234.355552 & -30.171385 & 3286 & 45837 & 0.22 & 0.63(2) & 0.97 \\
{[12]} & 113100013-441959935 & 4 & 172.750041 & -44.333297 & 1631 & 36669 & 0.99 & 0.02(2) & 0.96 \\
{[15]} & 081602164-530722970 & 4 & 124.009037 & -53.123042 & 4823 & 27502 & 0.87 & 0.34(3) & 0.95 \\
{[16]} & 175443398+214054818 & 3 & 268.680823 & 21.681869 & 1755 & 18335 & 0.45 &   & 0.95 \\
{[17]} & 065904044+162908685 & 3 & 104.766823 & 16.485772 & 5249 & 36669 & 1.47 & 0.14(2) & 0.94 \\
{[18]} & 182244519-541451730 & 4 & 275.685519 & -54.247701 & 5256 & 27502 & 1.22 & 0.11(4) & 0.94 \\
{[19]} & 054934271+051814610 & 3 & 87.392794 & 5.304042 & 2298 & 45837 & 0.43 &   & 0.93 \\
{[20]} & 075933618-173212537 & 3 & 119.890101 & -17.536806 & 1860 & 55004 & 1.23 &   & 0.93 \\
{[23]} & 181730853+272940139 & 3 & 274.378545 & 27.494468 & 1796 & 36669 & 1.79 &   & 0.91 \\
{[25]} & 024848742+191330571 & 3 & 42.203097 & 19.225140 & 1677 & 13751 & 0.30 & 0.06(2) & 0.88 \\
{[26]} & 201454150-302452196 & 3 & 303.725615 & -30.414491 & 2465 & 13751 & 0.48 & 0.32(2) & 0.88 \\
{[28]} & 201749047+620443509 & 3 & 304.454360 & 62.078774 & 916 & 36669 & 0.99 &   & 0.74 \\
{[30]} & 011559515+562506671 & 3 & 18.997963 & 56.418524 & 2756 & 45837 & 0.70 &   & 0.60 \\
\hline
\end{tabular}
\end{center}
\end{table*}

\begin{figure}
    \centering
    \includegraphics[scale=1.1]{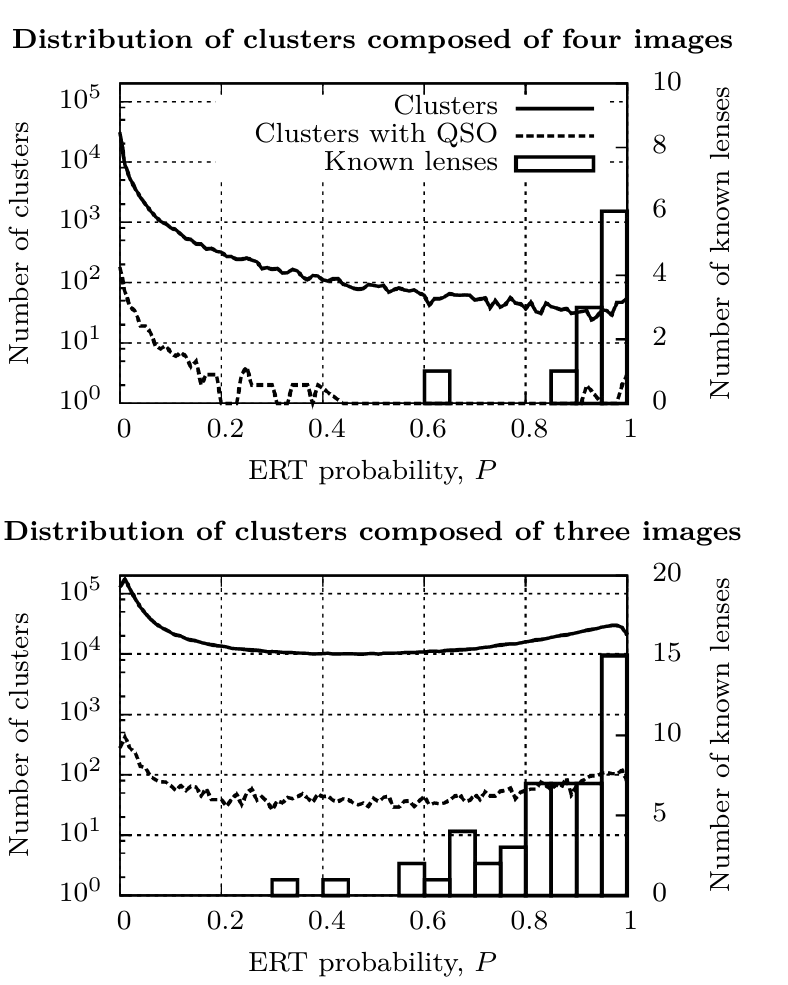}
    \caption{Distribution of the 70,697 clusters composed of four images and 2,058,962 clusters composed of three images extracted from the {\it Gaia} DR2 (see section \ref{sec:cluster_extraction}) with respect to their ERT probabilities (solid line). The distribution of the known lenses are represented as boxes whereas the distribution of the 6,944 clusters resulting from the cross-match we performed between our entire list of clusters and our compiled list of quasars is depicted as a dotted line in each graph (see Section \ref{subsubsec:gl3_candidates}). In cases where clusters are composed of three images, the ERT probability corresponds to the maximum of the ERT probabilities returned by the ABC, ABD, ACD and BCD models.}
    \label{fig:fnew}
\end{figure}

\begin{figure*}
\begin{center}
\includegraphics[width=\textwidth]{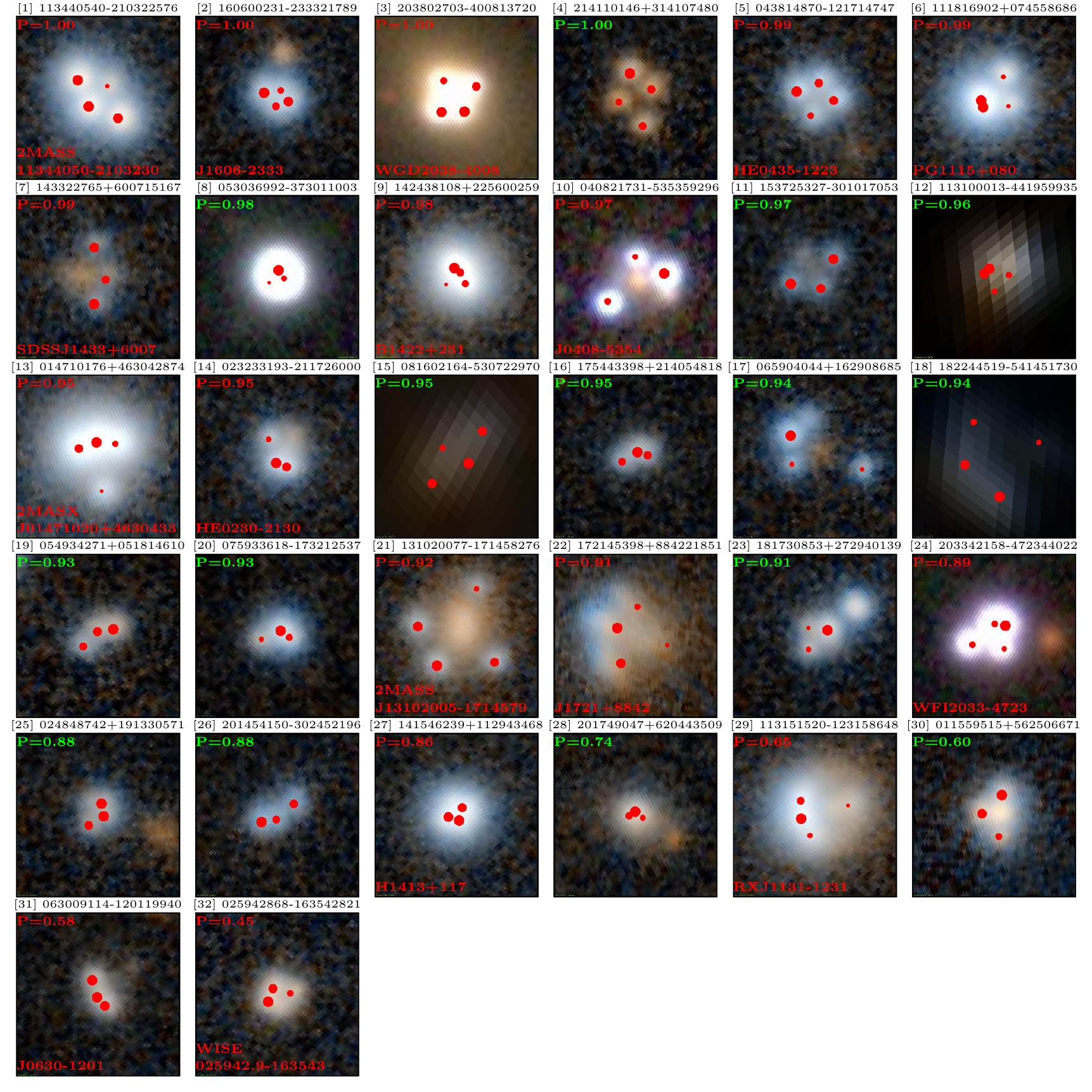}
\end{center}
\caption{Finding charts of the 17 known and 15 GL candidates that are contained in our catalogue of clusters (Appendix \ref{sec:catalogue}). These are ordered according to their ERT probabilities as provided in the upper left corner of each sub-plot. The common name of the known lenses is labelled in red in the lower-left corner of each sub-plot while the candidates we propose have their probabilities written in green fonts. Images [1], [2], [4-7], [9], [11], [13], [14], [16], [17], [19-23], [25-32] come from the Pan-STARRS survey \citep[][]{Chambers2016}, images [12], [15], [18] come from the Digitized Sky Survey II \citep{Lasker1996}, images [3], [8], [10], [24] come from the DES \citep{DES2016}. All images were collected from the ALADIN sky atlas \citep{Bonnarel2000} in a field of view of $10.8\arcsec \times 10.8\arcsec$ centered around the mean coordinates of the GL where east is to the left and north is up. Points are scaled according to the relative flux of the components with respect to the brightest image of each configuration.}
\label{fig:finding_charts}
\end{figure*}

\begin{figure*}
\begin{center}
\includegraphics[width=0.925\textwidth]{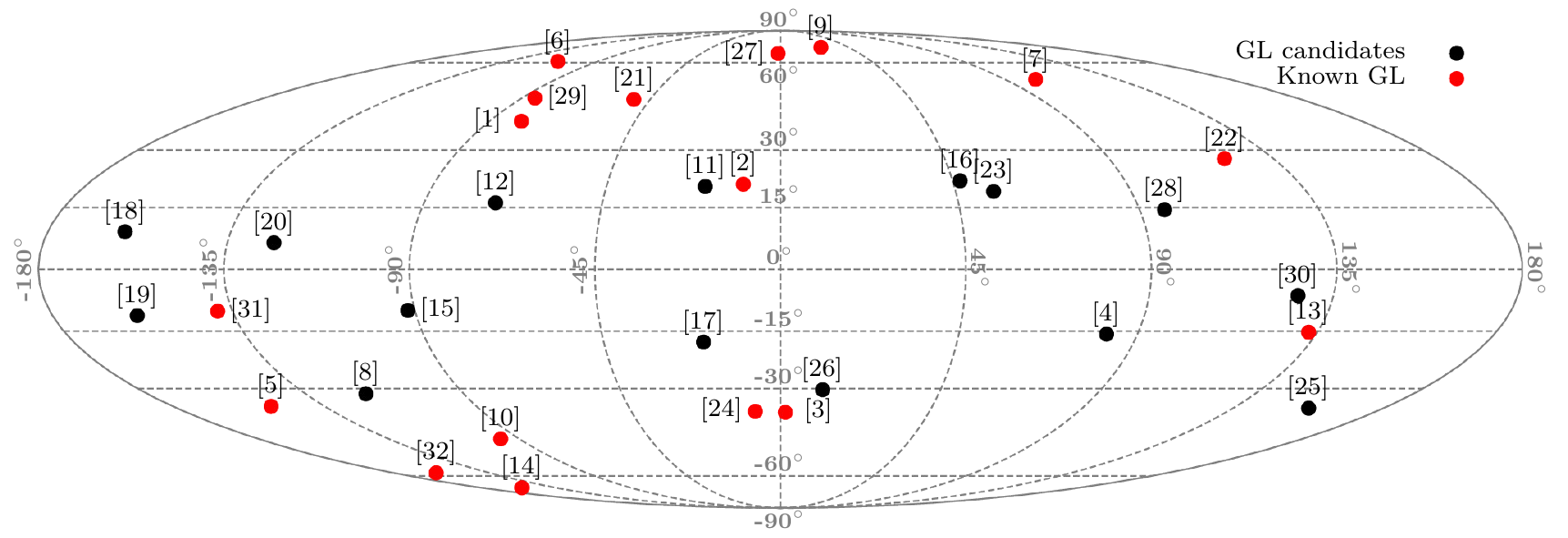}
\end{center}
\caption{Distribution of the known and new candidates GLs in Galactic coordinates. The numbers in square brackets refer to the candidate numbers from Table \ref{tbl:gl_promising} and Figure \ref{fig:finding_charts}.}
\label{fig:candidate_dist_gal}
\end{figure*}

    Finally, we applied the methodology developed in this work to the 2,129,659 clusters we extracted from the {\it Gaia} DR2 with a view of identifying GL candidates. The resulting catalogue of clusters, along with their associated ERT probabilities, is further described in Appendix \ref{sec:catalogue} though the distributions of these ERT probabilities regarding the clusters and the known lenses they contains are already provided in figure \ref{fig:fnew}. From this figure, we can see that most of the clusters have low ERT probabilities as, for example, 43.34\% of the clusters composed of three images and 89.52\% of the clusters composed of four images have ERT probabilities $P < 0.2$. Conversely, 10 out of the 11 known lenses composed of four {\it Gaia} detections and 36 out of the 50 known lenses composed of three detections\footnote{44 of the known gravitational lens systems composed of three images come from the combination of the images of the 11 known systems composed of four images while 6 are known systems for which an image was undetected (see Table \ref{tbl:known_gl} for details).} have an ERT probability $P > 0.8$. Note that the large number of clusters having high ERT probabilities amongst the clusters composed of three images is due to the choice we made to consider a single ERT probability that is taken as the maximum of the probabilities that are returned by the ABC, ABD, ACD and BCD models. This choice is justified by the fact that if we observed a genuine quadruply-imaged quasar having only three detections in {\it Gaia} DR2, we do not know which of the image was unobserved and consequently, in a conservative approach, we have to consider the ERT model that worked at best.
	
	In the following, we do not aim to provide an exhaustive list of the GLs that are contained in this catalogue, but rather to provide the community with a very pure list of GL candidates at the expense of a lower completeness. We should stress again here that GLs having prominent substructures in the lensing galaxy, multiple lensing galaxies and/or high variability will be hardly identified by the ERT as these are often not well modelled with NSIEg lens models.

\subsubsection{Systematic blind search of gravitational lenses composed of four images}
\label{subsubsec:gl4_candidates}

    In this first approach, we aimed at systematically and blindly identifying GL candidates that are composed of four {\it Gaia} detections while sharing common properties with the set of known lenses from Table \ref{tbl:known_gl}. The reason for considering these four image candidates apart from those composed of three images stands first in the benefit we can draw from the excellent performance of the ABCD ERT model. Furthermore, it is impossible to have a systematic search for lenses in the triplet regime because of both the relatively high FPR of the ABC, ABD, ACD and BCD ERT models (5\% $\lesssim$ FPR $\lesssim$ 20\%) and of the numerous clusters to which they will be applied (2,058,962 clusters composed of three images).
    
    According to the mean density of objects found around known GLs (see Table \ref{tbl:known_gl}) and based on the maximal absolute difference in color between their lensed images, $\Delta (G_{\mathrm{BP}}-G_{\mathrm{RP}})$, we first restricted our four image candidates to be situated in regions where $\rho < 3 \times 10^4$ objects deg$^{-2}$ while having a maximal absolute difference in color $\Delta (G_{\mathrm{BP}}-G_{\mathrm{RP}}) < 0.4$ mag, when available. Similarly, we also required our candidates to have an ERT probability of $P > 0.9$. Amongst the 10 clusters satisfying the aforementioned criteria, 7 are known lenses (2MASS J11344050-2103230, WGD2038-4008, HE0435-1223, PG1115+080, B1422+231, 2MASS J13102005-1714579 and J1721+8842) while three clusters (numbered [4], [15] and [18] in the finding charts from Figure \ref{fig:finding_charts} and Table \ref{tbl:gl_promising}) are new GL candidates. This first analysis already proved the identification capability of our approach based solely on data products from the {\it Gaia} DR2. Nevertheless, candidate number [4] is probably not a GL, to our opinion, mostly because of the red color of its constituent images. Similarly, the DSS2 images of candidates [15] and [18] do not have sufficient angular resolution for determining their lensing nature. Still, their relative image positions and fluxes are compatible with those produced by NSIEg-like lenses.

\subsubsection{Search for gravitational lenses around quasars and quasar candidates}
\label{subsubsec:gl3_candidates}
    
    A systematic blind search for new lensed systems where only three images are detected in {\it Gaia} is infeasible given the numbers we provided in figure \ref{fig:fnew}. Instead, constraints from external catalogues have to be used so as to lessen the number of candidate clusters. We know that lensed sources from GLs are always extragalactic objects, either active galactic nuclei (AGN) or galaxies. The latter are however extended objects of low surface brightness that will accordingly not be detected by {\it Gaia}. We thus performed a cross-match between our entire list of clusters, the compiled quasars list from \cite{KronMartin2018} and the C75 and R90 {\it WISE} AGN catalog from \cite{2018ApJS..234...23A}. The first of these lists consist of 3,112,975 candidate quasars compiled from the the million quasars catalogue \citep{Flesch2015,Flesch2017}, from a photometric selection of {\it WISE} AGN \citep{Secrest(2015)}, from the third release of the large quasar astrometric catalogue \citep{Souchay2015} and from the twelfth data release of the SDSS quasar catalogue \citep{Paris2017}. The R90 and C75 catalogues consist of 4,543,530 and 20,907,127 {\it WISE} AGN candidates, respectively, selected across the whole extragalactic sky based solely on mid-infrared colors. The R90 catalogue has a reliability of 90\%, while the C75 catalogue has a completeness of 75\%. 
    
    This cross-match resulted in 6944 clusters composed of three or four images for which at least one of them has a counterpart in our compiled list of quasars. Figure \ref{fig:fnew} gives the distributions of the ERT probabilities amongst these clusters. We note that these distributions are simply scaled versions of the distributions we got if no cross-match was performed, meaning that the vast majority of the clusters remains contaminants and not gravitational lens systems. Based on the same figure, we decided to set a threshold on the ERT probability of $P \geq 0.6$ which ensures that most of the known lenses will be detected, at the exception of four (over a total of 61 known systems: 11 with four {\it Gaia} detections, 44 combinations of three images from the latter and 6 having three {\it Gaia} detections). A visual inspection of the 2572 resulting clusters composed of more than two images then yielded the GL candidates numbered [8], [11], [12], [16], [17], [19], [20], [23], [25], [26], [28] and [30] from Figure \ref{fig:finding_charts} and Table \ref{tbl:gl_promising}. Despite the low cut we set on the ERT probability ($P \geq 0.6$), we may note that out of the twelve candidates we propose, ten have ERT probabilities higher or equal to 0.88, assessing the interest of the ERT for identifying GLs, even in the case where only three images are detected. 
    
    Finally, note that candidate number [12] was already present in \cite{KronMartin2018} and has since been spectroscopically confirmed as a GL \citep{2018arXiv181002624W}. Five other candidates were also spectroscopically confirmed through Keck/LRIS observations after the submission of this paper (candidates numbered [11], [17], [23], [25] and [26], Krone-Martins et al., in prep.). Three of these new GLs (numbered [11], [23] and [26]) were also independently confirmed by \cite{2018arXiv181004480L}. Candidate number [26] is however a doubly-imaged quasar that should hence be considered as a false positive from the ERT. On the other hand, the lensing nature of two candidates were denied (candidates numbered [16] and [30]) and led to inconclusive results regarding two others (candidates numbered [19] and [28]). The lensing nature of other candidates currently remains uncertain but plausible.
    

\section{Conclusions}
\label{sec:conclusion}

	In this work, we aimed to discover quadruply-imaged lens candidates through the use of a supervised learning method, called Extremely Randomized Trees (ERT), applied over the whole {\it Gaia} DR2. To train ERT, we simulated the relative positions and flux ratios of 106,623,188 quadruply-imaged systems based on a non-singular isothermal ellipsoid lens model in the presence of external shear. The performance of our ERT models were probed using both simulations and real observations from {\it Gaia} DR2. From known quadruply-imaged systems having all components detected in {\it Gaia} DR2, 12 out of 13 are successfully recovered by our method along with a misclassification rate of fortuitous clusters of stars as lens that is below one percent. Similarly, 92.5\% of the simulated lens systems are identified with the same misclassification rate.
    
    The performance of our ERT models in identifying quadruply-imaged systems where only three components are detected in {\it Gaia} DR2 are evaluated by removing one image from each of the simulated and known quadruply-imaged systems. This resulted in the correct identification of 10 out of 13 known lensed systems with a misclassification rate below 7.5\% once the two brightest images of the lens are observed together and of $\sim 20$\% otherwise. For the same identification rate, tests performed on simulations provided a similar misclassification rate of 7.74\% for configurations where the two brightest images are present and of $\sim 10$\% otherwise.
    
    We applied our ERT models to 70,697 clusters composed of four images and to 2,058,962 clusters composed of three images we extracted out of the {\it Gaia} DR2. Beforehand, a filtering of the {\it Gaia} DR2 sources was use in order to remove non-stationary objects based on the parallax and proper motions of each source. Clusters were also restricted to have a maximal separation between images of $6\arcsec$, a maximal absolute difference of $G$ magnitude between components below $4$ mag while standing in regions of the sky where the mean field density is below $6 \times 10^4$ objects deg$^{-2}$.
    
    Benefiting from the high identification rate of our ERT model in cases where all four components from quadruply-imaged systems are detected and of the low associated misclassification rate of clusters of stars as gravitational lens, we succeeded in isolating seven known gravitational lenses composed of four images based on simple cuts in the mean field density, in the maximal absolute difference in color and in the discriminant value provided by the ERT model. Three clusters are also retrieved through this straight selection and are hence promoted as plausible lens candidates.
    
    In addition to this {\it Gaia}-only approach, we performed a cross match between our list of clusters from {\it Gaia} DR2 sources and compiled lists of spectroscopically confirmed quasars and photometric quasar candidates. We visually inspected the resulting 2572 clusters for which the ERT models predicted a reasonably good agreement between these clusters and the relative positions and flux ratios from a non-singular isothermal ellipsoid lens model in the presence of an external shear. In total, 15 new lens candidates were identified, including one quadruply-imaged system that was recently spectroscopically confirmed.

	The present method succeeded in finding highly probable quadruply-imaged quasar candidates out of which five were recently spectroscopically confirmed. The low number of lens candidates we identified from {\it Gaia} data, with respect to \citet{Finet2016} predictions, can mostly be explained by the fact that the majority of gravitational lenses present in {\it Gaia} DR2 have less than three lensed images published in the catalogue, as shown by \citet{Ducourant2018}. We expect that {\it Gaia} DR3 and later DR4 will improve on this, due to a less stringent filtering of the published sources and improved processing. Meanwhile, the present method can be used on other catalogues, as it solely relies on astrometric and photometric data. Applications are already foreseen for the upcoming {\it Gaia} DR3 and combinations of already available catalogues (e.g. PanSTARRS, DES, SDSS and {\it Gaia} DR2).

\begin{acknowledgements}
      LD and JS acknowledge support from the ESA PRODEX Programme `{\it Gaia}-DPAC QSOs' and from the Belgian Federal Science Policy Office.
       AKM acknowledges the support from the Portuguese Funda\c c\~ao para a Ci\^encia e a Tecnologia (FCT) through grants SFRH/BPD/74697/2010, PTDC/FIS-AST/31546/2017, from the Portuguese Strategic Programme UID/FIS/00099/2013 for CENTRA, from the ESA contract AO/1-7836/14/NL/HB and from the Caltech Division of Physics, Mathematics and Astronomy for hosting a research leave during 2017-2018, when this paper was prepared. 
      OW is supported by the Humboldt Research Fellowship for Postdoctoral Researchers.
      SGD and MJG acknowledge a partial support from the NSF grants AST-1413600 and AST-1518308, and the NASA grant 16-ADAP16-0232.
      The work of DS was carried out at the Jet Propulsion Laboratory,
California Institute of Technology, under a contract with NASA.
      We acknowledge partial support from `Actions sur projet INSU-PNGRAM', and from the Brazil-France exchange programmes Funda\c c\~ao de Amparo \`a Pesquisa do Estado de S\~ao Paulo (FAPESP) and Coordena\c c\~ao de Aperfei\c coamento de Pessoal de N\'ivel Superior (CAPES) -- Comit\'e Fran\c cais d'\'Evaluation de la Coop\'eration Universitaire et Scientifique avec le Br\'esil (COFECUB).
      This work has made use of the computing facilities of the Laboratory of Astroinformatics (IAG/USP, NAT/Unicsul), whose purchase was made possible by the Brazilian agency FAPESP (grant 2009/54006-4) and the INCT-A, and we thank the entire LAi team, specially Carlos Paladini, Ulisses Manzo Castello, Luis Ricardo Manrique and Alex Carciofi for the support.
      This work has made use of results from the ESA space mission {\it Gaia}, the data from which were processed by the {\it Gaia} Data Processing and Analysis Consortium (DPAC). Funding for the DPAC has been provided by national institutions, in particular the institutions participating in the {\it Gaia} Multilateral Agreement. The {\it Gaia} mission website is:
http://www.cosmos.esa.int/gaia. Some of the authors are members of the {\it Gaia} Data Processing and Analysis Consortium (DPAC).
\end{acknowledgements}

\bibliographystyle{aa}
\bibliography{bibliography}

\begin{appendix}
\section{The non-singular isothermal ellipsoid lens model in the presence of an external shear}
\label{sec:nsieg}
The non-singular isothermal ellipsoid lens model in the presence of an external shear (NSIEg) is characterized by $\kappa$, the {\it dimensionless surface mass density} projected in the lens plane and defined by
\begin{equation}
\kappa(x,y) = \frac{b}{2} \left( s^2 + x^2 + \frac{y^2}{q^2} \right)^{-\frac{1}{2}} \ ,
\label{eq:nsieg_kappa}
\end{equation}
where the coordinates $(x,y)$ locate a position in the lens plane, $s$ corresponds to the deflector core radius, $q$ is the ratio of the minor to the major axes of the elliptical isodensity contours, and $b$ is considered here as a normalizing factor. From \cite{Keeton1998}, the two components of the corresponding {\it scaled deflection angle}, $\vec{\alpha}$, are respectively given by
\begin{equation}
\alpha_x(x,y) = \frac{b\,q}{e}\, \tan^{-1}\left(\frac{e\,x}{\psi + s}\right) \ ,
\label{eq:nsieg_alphax}
\end{equation}
and
\begin{equation}
\alpha_y(x,y) = \frac{b\,q}{e}\, \tanh^{-1}\left(\frac{e\,y}{\psi + q^2 s}\right) \ ,
\label{eq:nsieg_alphay}
\end{equation}
where $e = \sqrt{1-q^2}$ is defined as the {\it eccentricity} of the isodensity contours and $\psi^2 = q^2 \left( s^2 + x^2 \right) + y^2$. The contribution of more distant massive objects to the deflection is taken into account through an {\it external shear} term of the form
\begin{equation}
\vec{\alpha}_{\gamma}(x,y) = \gamma \,
\begin{pmatrix}
\;\cos 2\omega & \sin 2\omega \; \\
\;\sin 2\omega & -\cos 2\omega \;
\end{pmatrix}
\begin{pmatrix}
\;x\; \\
\;y\;
\end{pmatrix} \ ,
\end{equation}
where $\gamma$ is the {\it shear intensity} and $\omega$ its orientation. Finally the position $\vec{\theta_s} = \left( x_s, y_s \right)$ of a point-like source is related to a lensed image position $\vec{\theta} = \left( x, y \right)$ through the so-called {\it lens equation}
\begin{equation}
\vec{\theta_s} = \vec{\theta} - \vec{\alpha} - \vec{\alpha}_{\gamma} \ ,
\label{eq:nsieg_lens}
\end{equation}
and the associated {\it magnification factor} $\mu(\vec{\theta})$ is then defined by
\begin{equation}
\mu(\vec{\theta}) = \left|\, \det \mat{A}(\vec{\theta})\, \right|^{-1} \ ,
\label{eq:nsieg_mu}
\end{equation}
where the Jacobian matrix $\mat{A}(\vec{\theta}) = \partial \vec{\theta_s}(\vec{\theta}) / \partial \vec{\theta}$ is called the {\it amplification matrix}.

\section{The Gaia GraL catalogue of clusters from {\it Gaia} DR2 sources}
\label{sec:catalogue}
\begin{table*}
\caption{Description of the fields contained in the {\it Gaia} GraL catalogue of clusters from {\it Gaia} DR2 sources.}
\label{tbl:catalogue_field_description}
\footnotesize
\begin{center}
\small
\begin{tabular}{rrcp{12cm}}
\hline
\hline
Num. & Field name & Unit & Description \\
\hline
1. & {\tt row} & & Unique identifier of the row in the catalogue. Each combination of the {\tt name} and of the {\tt SourceId} is always associated with a unique {\tt row}. \\
\hline
2. & {\tt name} & & Unique identifier of the cluster. This identifier is based on the mean position of the images of the cluster taken in sexagesimal coordinates. \\
3. & {\tt nimg} & & The number of images constituting the cluster ({\tt nimg} $\in \left\lbrace 3, 4\right\rbrace$). \\
4. & {\tt size} & [$\arcsec$] & The maximal angular separation between any two images of the cluster ({\tt size} $\leq 6\arcsec$). \\
5. & {\tt known} & & Identifier in the list of known and candidate GL from \cite{Ducourant2018}. Empty if this cluster is not recognized as a known GL or candidate. \\ 
6. & {\tt density} & [obj. deg$^{-2}$]  & The mean density of objects surrounding the cluster, estimated by counting the {\it Gaia} DR2 objects falling in a $30\arcsec$ radius around one of its constituent members ({\tt density} $ \leq 6 \times 10^4$ objects deg$^{-2}$).\\
7. & {\tt dmag} & [mag] & The maximal absolute difference in the $G$ magnitude, $\Delta G$, between the images of the cluster ({\tt dmag} $ \leq 4$ mag). \\
8. & {\tt dcolor} & [mag] & The maximal absolute difference in color, $\Delta (G_{\mathrm{BP}}-G_{\mathrm{RP}})$, between the images of the cluster. Empty if less than two images comes along with color informations (i.e. {\tt BPmag} or {\tt RPmag} is empty).\\
9. & {\tt ncolor} & & The number of images having color information that are used in the computation of {\tt dcolor}.\\
10. & {\tt P} & & ERT probability associated with this cluster (see Section \ref{sec:method}). \\
 & & & If {\tt nimg} = 4, then this parameter corresponds to the ERT probabilities computed based on the ABCD model. \\
 & & & If {\tt nimg} = 3, then this field is taken as the maximum of {\tt PABC}, {\tt PABD}, {\tt PACD} and {\tt PBCD}. \\
11. & {\tt PABC} & & ERT probability computed based on the ABC model. Empty if {\tt nimg} = 4. \\
12. & {\tt PABD} & & ERT probability computed based on the ABD model. Empty if {\tt nimg} = 4\\
13. & {\tt PACD} & & ERT probability computed based on the ACD model. Empty if {\tt nimg} = 4\\
14. & {\tt PBCD} & & ERT probability computed based on the BCD model. Empty if {\tt nimg} = 4\\
15. & {\tt rev} & & Revision number in case the ERT probabilities are to be re-evaluated (e.g. because more accurate models are available) or if this cluster was since spectroscopically confirmed as a gravitational lens system. Initially set to 0.\\
\hline
16. & {\tt SourceId} & & Unique source identifier from {\it Gaia} DR2. \\
17. & {\tt RA} & [$\degr$] & Right ascension of the source in ICRS coordinates from {\it Gaia} DR2. \\
18. & {\tt DEC} & [$\degr$] & Declination of the source in ICRS coordinates from {\it Gaia} DR2. \\
19. & {\tt Gmag} & [mag] & $G$-band mean magnitude from {\it Gaia} DR2. \\
20. & {\tt BPmag} & [mag] & Integrated BP mean magnitude from {\it Gaia} DR2. \\
& & & Empty if no BP magnitude is available. \\
21. & {\tt RPmag} & [mag] & Integrated RP mean magnitude from {\it Gaia} DR2.\\
& & & Empty if no RP magnitude is available, this field is set to `?'. \\
\hline
\end{tabular}
\end{center}
\end{table*}
	The {\it Gaia} GraL catalogue of clusters from {\it Gaia} DR2 sources can be retrieved in electronic form using the CDS facilities at \href{ftp://cdsarc.u-strasbg.fr/}{ftp://cdsarc.u-strasbg.fr/}\textcolor{red}{XXX} or via \href{http://cdsweb.u-strasbg.fr/cgi-bin/gcat?J/A+A/}{http://cdsweb.u-strasbg.fr/cgi-bin/gcat?J/A+A/}\textcolor{red}{XXX}. The catalogue is composed of all 2,129,659 clusters identified in Section \ref{sec:cluster_extraction} along with the ERT probabilities computed for each of them (see Section \ref{sec:method}). For ease of identifying the images constituting each cluster as well as for facilitating the cross-match against other catalogues based on the individual components of the clusters, each entry from the catalogue corresponds to a single {\it Gaia} DR2 source within the cluster. Consequently clusters composed of three or four images have, respectively, three and four associated entries in the catalogue along with the fields that are common to the cluster they belong to. The fields constituting each row of the catalogue are detailed in Table \ref{tbl:catalogue_field_description}.

Our objective here is not to provide a list of GL candidates, as we do in Sections \ref{subsubsec:gl4_candidates} and \ref{subsubsec:gl3_candidates}, but to provide the user with a catalogue where (s)he can easily get hints on the lensing nature of some of their observational targets, at least regarding GLs that are reproducible through a NSIEg lens model. This approach also justifies the inclusion of clusters standing in regions of the sky we know to be densely populated and where the contamination rate of GL candidates will be typically very high. The removal of the clusters having a field density higher than $\rho > 3 \times 10^4$ objects deg$^{-2}$ effectively reduces their number by a factor of ten (205,004 remaining clusters). 

	To our knowledge, the present catalogue is the first one to provide a discriminating value associated with each cluster that reflects the ability for a given GL model to reproduce the observed configuration of images. These ERT probabilities can provide a straight binary classification as, for example, 96.31 per cent of the four image configurations have $P < 0.5$ whereas 86.11 per cent of the three image configurations have $P < 0.9$. The threshold we set on $P$ are obviously application-dependent and should be set in agreement with the ROC curves we described in Section \ref{sec:method}. Finally, in a conservative approach, we do not set any cut on the difference in color between images of the clusters, $\Delta (G_{\mathrm{BP}}-G_{\mathrm{RP}})$. Whenever available, these however provide an important criterion for identifying GLs as we do not expect the colors to vary much between the images of GLs (see Sect. \ref{subsec:gl_candidates}, for examples).
\end{appendix}

\end{document}